\documentstyle[10pt,aas2pp4,amssym]{article}   

\accepted{September 8 - 1997}

\journalid{493}{February 1 - 1998}

\lefthead{G. Giovannini et al.}
\righthead{Paper VIII: 3C338}

\begin{document}

\title
{ VLBI Observations of a Complete Sample of Radio Galaxies \\
VIII - Proper Motion in 3C338}

\author{G. Giovannini\altaffilmark{1}}
\affil{Istituto di Radioastronomia, via Gobetti 101, 40129 Bologna, ITALY}
\authoremail{ggiovannini@astbo1.bo.cnr.it}
\author{W.D. Cotton}
\affil{N.R.A.O., 520 Edgemont Rd,  Charlottesville VA 22903-2475, USA}
\authoremail{bcotton@nrao.edu}
\author{L. Feretti}
\affil{Istituto di Radioastronomia, via Gobetti 101, 40129 Bologna, ITALY}
\authoremail{lferetti@astbo1.bo.cnr.it}
\author{L. Lara}
\affil{Instituto de Astrofisica de Andalucia, CSIC, Apartado 3004, 18080,
Granada, SPAIN }
\authoremail{lucas@iaa.es}
\and
\author{T.Venturi}
\affil{Istituto di Radioastronomia, via Gobetti 101, 40129 Bologna, ITALY}
\authoremail{tventuri@astbo1.bo.cnr.it}

\altaffiltext{1}{Dipartimento di Astronomia, via Zamboni 33, 40126 
                 Bologna, ITALY}
\vskip 1truecm
\centerline{===========================================================}

\centerline{\bf Astrophysical Journal Vol. 493 - Feb. 1 1998 - in press}

\centerline{preprint n.: BAP 09-1997-027 IRA;  available in: 
            http://boas3.bo.astro.it/bap/BAPhome.html}

\centerline{===========================================================}

\begin{abstract}
We present new VLA, MERLIN and VLBI images for the 
radio galaxy 3C 338, and the results of the monitoring of its
arcsecond core flux density.
Present high sensitivity observations allow us to investigate 
the radio structure of this source and to confirm the presence of two
symmetric parsec scale jets. Morphological changes between different epochs 
are evident and a proper motion with $\beta \sim$ 0.4h$^{-1}$ has been 
derived allowing us to give a lower limit constraint for the Hubble
constant. While the steep spectrum large scale structure of 3C 338 could be  a
relic emission, the small scale structure looks young, 
similar to the high power MSOs found at high redshift.
\end{abstract}


\singlespace
\section{Introduction}
The radio galaxy 3C338 is classified as a FR I radio source (see Fanaroff and 
Riley, 1974 for the FR I definition) and shows central optical [OIII] line
emission (Fisher et al., 1995). It has the steepest radio
spectrum at 
cm wavelength of any 3CR source except 3C318.1 (Feretti et al. 1993). It is
associated with the multiple nuclei cD galaxy NGC6166 at the
center of the cooling flow cluster of galaxies A2199. Lucey et al. (1991)
found for this cluster a peculiar velocity not significantly different from
zero, therefore we derive its distance from the measured cluster redshift
z = 0.03023 (Zabludoff et al., 1993).

Short exposure plates by Minkowski (1961) and Burbidge (1962) showed that the
cD core consists of 4 separate optical components, with the brightest one
coincident with the radio core. This complex structure is confirmed 
in a recent HST image (Capetti private communication).

In the low resolution radio maps, 3C338 shows a total extension of $\sim$ 2' 
with a
core emission and 2 symmetric lobes slightly misaligned with respect to the
core. High resolution radio maps (Burns et al., 1983) reveal the presence of a
peculiar jet like filament within the extended structure of 3C338, but
detached and significantly offset to the south of the compact radio core. This
feature has a very steep spectrum as the entire diffuse structure does. 

Burns et al. (1983) suggested two possible explanations for this peculiar 
structure:
a) the ram pressure of a highly asymmetric cooling flow onto the cD galaxy;
b) the motion of the radio core within the cD galaxy. If the radio core
stopped its activity at some time and moved from its position, it could 
have left a steep spectrum aged radio jet behind.
More recent observations (Ge et al. 1994) also 
identified a weak radio emission
from the second optical nucleus of NGC 6166 and revealed faint symmetric 
jets on both sides of the 3C338 core, with a size of about 15".
These new observations show also high RM values, ranging  from -2000 to +2000 
rad m$^{-2}$ in the two extended lobes, in the jet-like filament and also in
the east jet. No polarized emission was detected from the core. 
The variation of 
polarization angles imply the existence of a screen in front of the radio 
source. A strong, ordered magnetic field with a dense medium in the center of 
a cooling flow could account for the observed results.
In a recent paper Owen et al., (1997) using new ROSAT HRI X-ray data showed
that the conditions in the central regions of 3C338 are very complex. 
The strong cooling
flow is not symmetric. They suggest that the conditions in the central cooling 
core could have disrupted the radio jet and created the low 
brightness emission.
In this scenario, the bright emission south of the core is a high-emissivity
transient filament, currently overpressured, for instance due to strong
turbulence in the region. The short-scale radio jets found by Ge et al. (1994) 
would be a new, young radio emission from 3C338 which is now in a radio-active
phase.

At parsec resolution, this source was first
mapped by Feretti et al. (1993), who
also pointed out that the arcsecond core radio emission
is strongly variable. They obtained a
high resolution VLBI map showing a central dominant core emission with a
symmetric two-sided jet structure. This structure is aligned within 
5$^{\circ}$ of
the arcsecond jet structure found by Ge et al. (1994) on both sides of the
arcsecond core. The parsec-scale symmetric structure can be formed by jets 
either moving at high velocity in the plane of the sky or with a 
non-relativistic speed. 
  
In this work, we present new data obtained with the Very Large Array (VLA), 
the Multi-Element Radio Linked Interferometer Network (MERLIN) and the Very
Long Baseline Interferometer (VLBI), for a detailed study
of this peculiar source and a discussion of its core radio properties.
We use a Hubble constant H$_0$ = 100 km sec$^{-1}$ Mpc$^{-1}$
therefore at the  distance of 3C338, 1 mas corresponds to 0.41 pc.

\section{Observational Data}

\subsection{VLA Data}

\subsubsection{Monitoring of the arcsecond core flux density}

We observed the arcsecond core radio emission of this source with the VLA
in the A-array configuration at different frequencies in order to derive
the arcsecond core flux density. The source was observed for about 10 minutes
at each frequency at different hour angles to obtain a better uv-coverage. The
data have been calibrated in the standard way using the Astronomical Image
Process System (AIPS) and
have been reduced using the MX or IMAGR AIPS tasks. The core flux density
was obtained by fitting an elliptical Gaussian to the nuclear source (IMFIT).
Our data set includes also VLA data when the VLA was used as a phased array
during VLBI observations.
In addition, literature data and unpublished VLA
archive data were used. The flux density monitoring data are shown in Table 
\ref{t1} and in Fig. 1. The arcsecond core source has shown 
2 main flares about 15 years
apart and is now in a low state. Unfortunately, we have no high frequency
data in the time range
1980 - 1990 therefore we cannot exclude the possibility that the core has had
more flares in this period.
Comparing data at the same epoch, the spectral index of the arcsecond core
turns out to be $\alpha$ $\sim$ 0.3 between 1.4 and 8.4 GHz 
(S($\nu) \propto \nu^{-\alpha}$).

\begin{table}
\caption{Arcsecond core flux densities \label{t1}}
\begin{flushleft}
\begin{tabular}{lllll}
\tableline
Epoch   & S$_{1.4 GHz}$ & S$_{5 GHz}$ & S$_{8.4 GHz}$ &  Ref. \\
Mon.Year & mJy          & mJy         & mJy           & \\
\tableline
Dec.   1974 &  --  &  164  &  --    &   5  \\
Feb.   1975 &  --  &  150  &  --    &   5  \\
Dec.   1975 &  --  &  150  &  --    &   5  \\
Aug.   1976 &  --  &  149  &  --    &   5  \\
Jan.   1977 &  --  &  149  &  --    &   5  \\
May    1978 &  --  &  130  &  --    &   5  \\
Feb.   1980 &  --  &  115  &  --    &   5  \\
May    1980 & 153  &  105  &  --    &   1  \\
Sep.   1982 & 159  &   --  &  --    &   2  \\
Dec.   1984 & 146  &   --  &  --    &   3  \\
Apr.   1989 &  --  &  154  &  --    &   6  \\
Apr.   1990 &  --  &  168  &   143  &   7  \\
Mar.   1991 &  --  &   --  &   136  &   4  \\
June   1991 & 181  &   --  &   --   &   4  \\
Aug.   1991 &  --  &  158  &   127  &   4  \\
Jan.   1993 & 184  &  140  &   --   &   4  \\
Apr.   1994 & 167  &  111  &   93   &   4  \\
Nov.   1994 &  --  &  113  &   87   &   4  \\
Sep.   1995 &  --  &  111  &   87   &   4  \\
Oct.   1995 &  163 & --    &   --   &   4  \\
\tableline
\label{t1}
\end{tabular}
\end{flushleft}
\tablerefs{
1: Burns et al., 1983; 2: Parma et al., 
\goodbreak
\parindent 0pt
1986; 3: de Ruiter et al., 1986; 4: present paper; 
\goodbreak
\parindent 0pt
5: Ekers et al., 1983; 6: Ge p.c.; 7: Feretti et al., 1993.
}
\end{table}

\begin{figure}
\figurenum{1}
\plotfiddle{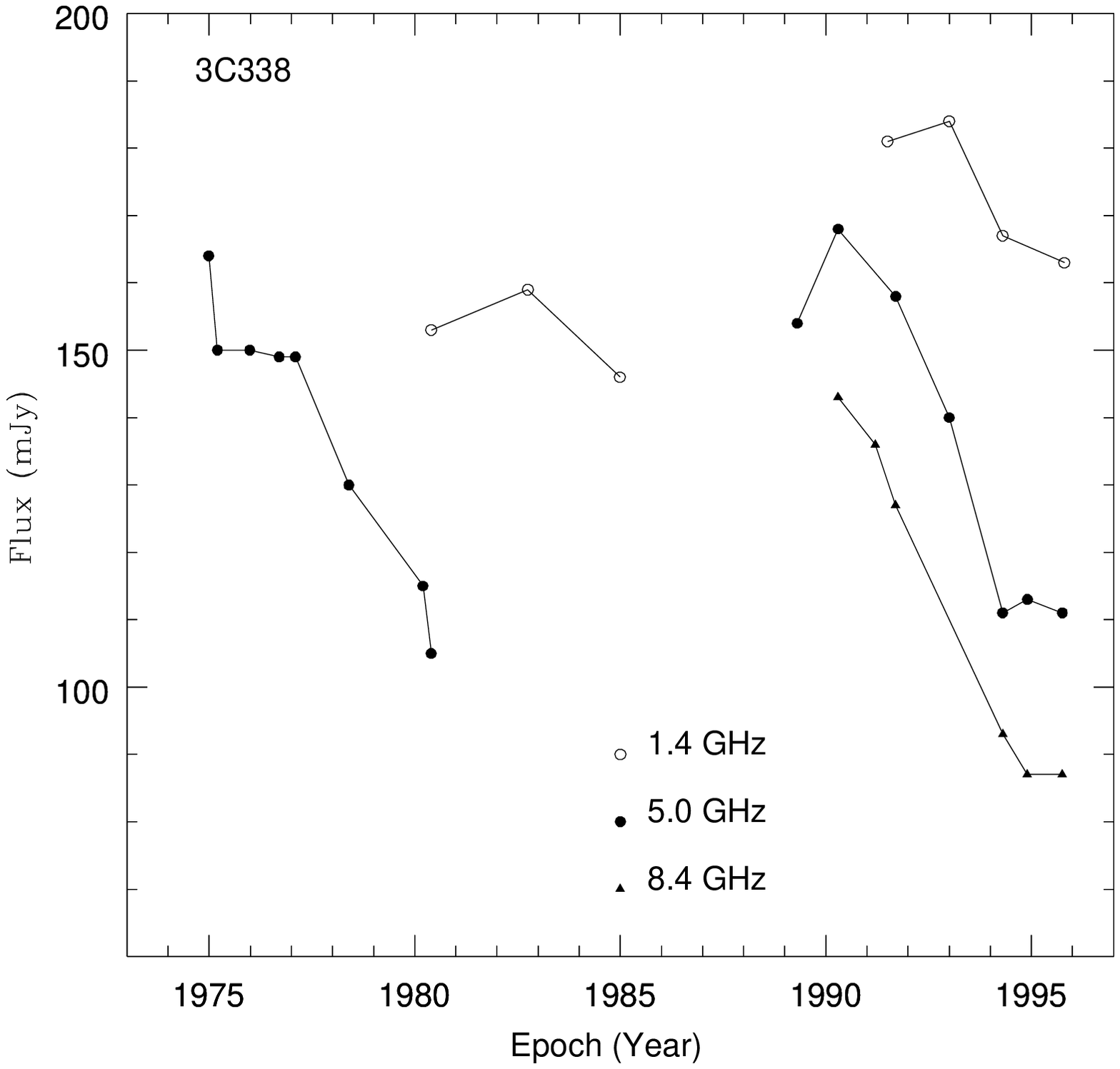}{7truecm}{0}{50}{50}{-144}{-72}
\caption{
Flux density measures of the 3C338 arc-second core. Connection
lines are for display use only and not a best fit. References to the different
measures are given in Table 1. }
\end{figure}

\subsubsection{VLA maps}

High quality images of 3C338 at the arcsecond resolution were obtained from the
long VLA observations made as part of the VLBI sessions as a phased
array.
In Fig. 2a we present the VLA map obtained at 1.7 GHz with the
VLA in the VLBI session on 1991 June 18. 
It was deconvolved using the Maximum Entropy Method (AIPS
task VTESS) which shows in better detail the low brightness extended regions.
The nuclear source has been partially subtracted. The central region is easily
visible with two symmetric, short jets which end in faint hot spots. The
radio structure of this small region is similar to that of extended
FR II radio galaxies but on a much smaller scale. 

\begin{figure}
\figurenum{2}
\plotfiddle{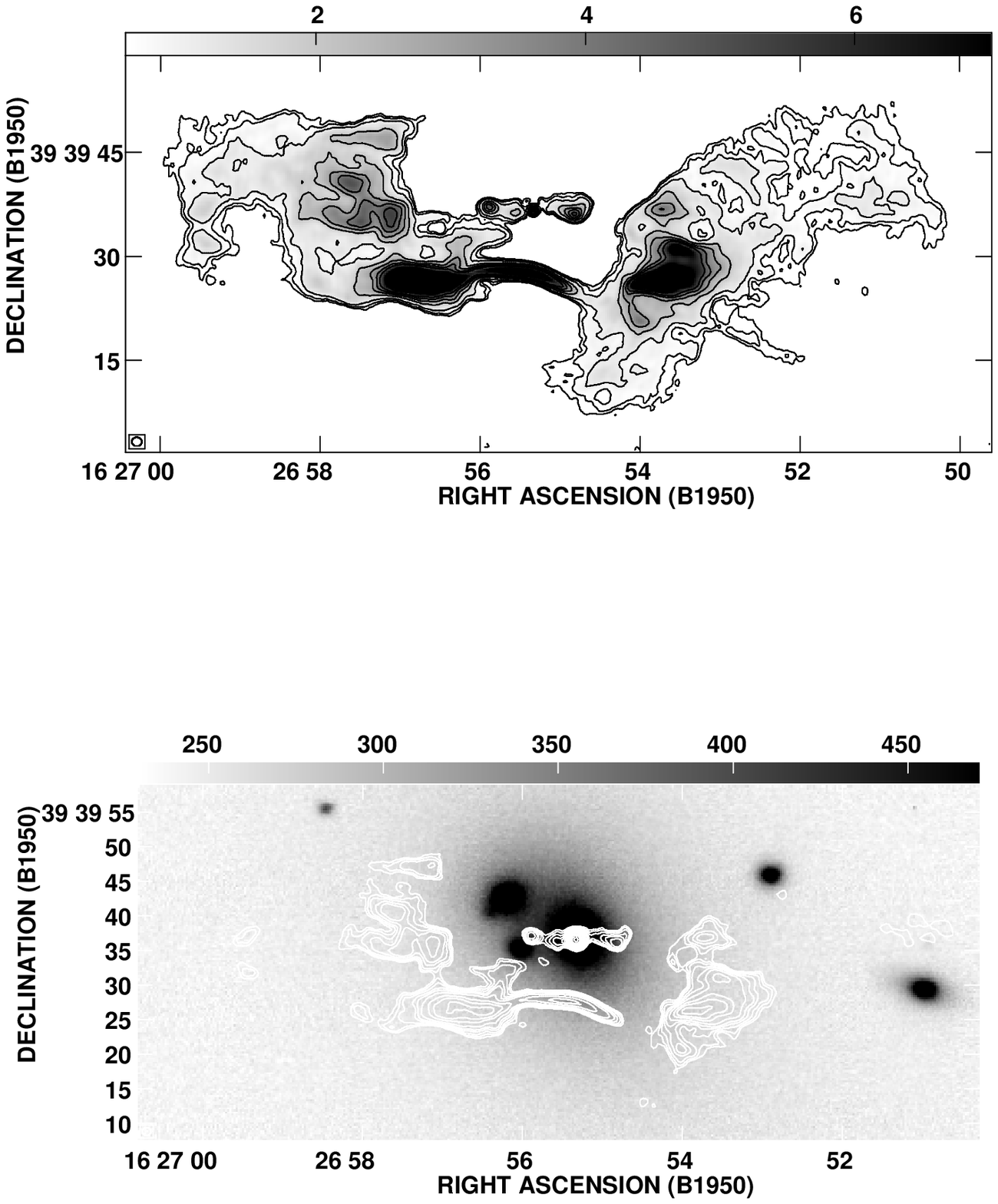}{10truecm}{0}{60}{60}{-180}{-144}
\caption{a- VLA map of 3C338 at 1.7 GHz deconvolved with
the Maximum Entropy Method. The nuclear source has been partially subtracted.
The HPBW is 1.42'' $\times$ 1.28'' in PA 84$^{\circ}$ and the noise level is 
0.08 mJy/beam.
Contour levels are: 0.4 0.6 1 2 3 4 5 7 10 mJy/beam.
b- Optical CCD map of 3C338 (grey levels) super-imposed with the
1.7 GHz radio map (contour levels). }
\end{figure}

The extended low brightness region as well as the peculiar jet--like
filament structure are evident in Fig. 2a. The filament structure is visible
beyond the west lobe.
In Fig. 2b, we have superimposed the radio map onto the optical image
taken from a 
CCD frame in the V band made with the 2.1 m reflector at San Pedro Martir (Baja
California) belonging to the Osservatorio Astronomico Nacional de Mexico, and
kindly provided by G. Gavazzi (see Gavazzi et al. 1995
for details). The optical image shows the three main optical nuclei
of 3C338 (the fourth, being very weak, is barely visible as a small extension
to the east of the northern one). The radio emission is 
identified with the dominant optical nucleus which is more diffuse, but
with a bright unresolved optical emission coincident with
the radio core. The two secondary nuclei are more compact with a
bright central emission.

\begin{figure}
\figurenum{3}
\plotfiddle{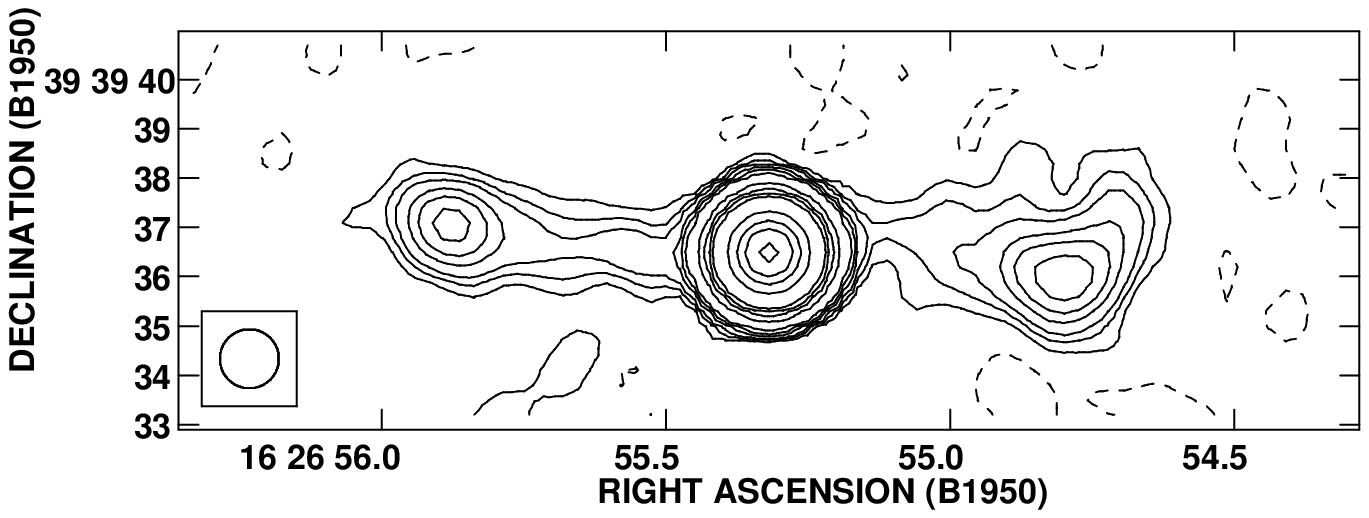}{3.5truecm}{0}{55}{55}{-170}{-190}
\caption{
Isocontour map of the central region of 3C338 obtained with
the VLA at 5.0 GHz. The HPBW is 1.2'' and the noise level is 0.04 mJy/beam.
Contour levels are: -0.1 0.1 0.2 0.3 0.5 0.7 1 3 5 7 10 30 50 70 100
mJy/beam.}
\end{figure}

In Fig. 3 we show the core region at 5.0 GHz from VLA data obtained in the
VLBI session on
1995 September 11. The VLA was in the A/B configuration with most of the 
telescopes
in the B configuration; therefore, the resolution of this map is very
similar to that
presented in Fig. 2. The core source is the dominant feature of this
small size structure. Two barely resolved symmetric jets connect the core 
to the two hot spots, where the radio structure seems to terminate. 
Both hot spots are resolved and their deconvolved size is 0.8'' (0.33 Kpc -
East one) and 1.4'' (0.57 Kpc - West one).
No connection between this central radio
emission and the extended steep spectrum low brightness emission is visible.
The distance between the core and the eastern and western hot spots is 6.4" 
(3.0 kpc) and 6.0" (2.8 kpc), respectively. The properties and shape of this
structure appear to resemble Compact Symmetric Sources (see Sect. 3.4).

We confirm the existence of the faint radio emission from the second optical 
nucleus (Ge et al. 1994) with a flux density of $\sim$ 0.5 mJy at 5 GHz 
(figure not presented here).
We did not attempt to study the polarized emission since we do not have the
frequency coverage necessary to study this 
high RM source, and refer to Ge et al.
(1994) for a detailed polarization study of 3C338.

We used the 1.7 and 5.0 GHz data to derive a spectral index map of the
extended structure of 3C338 at arcsecond
resolution (not shown here), the uv-coverage of the two datasets being very
similar.
The structure in the central region has a moderately steep spectrum
($\alpha$ = 0.7 - 1.5 in the two symmetric jets; $\alpha$ = 1.1 in the two
hot spots), while the detached extended lobes and the
relic jet structure have a very steep spectrum ($\alpha$ = 1.7 - 3.0), in 
agreement with Burns, (1983).

\subsection{MERLIN Data}

We observed 3C338 with the MERLIN array on 1995 October 29 at 1.66 GHz with
a 15 MHz bandwidth for 12 hours. We used the following telescopes: Defford,
Cambridge, Knockin, Wardle, Darnhall, MK2, Tabley. The data were edited and
amplitude calibrated in Jodrell Bank using the standard procedure based on
the OLAF programs. 3C286 was used as amplitude calibrator. The data were then
written in FITS format and loaded into AIPS where the phase calibration using
standard MERLIN phase calibrators was carried out. The source was then
mapped and the data self-calibrated following the standard procedure. In Fig. 4
we present the MERLIN map obtained using {\it Natural} weighting.
The large scale structure is
completely resolved and the filament jet--like structure to the south of
the core shows a
uniform distribution with no evidence of unresolved knots inside. The core is
easily visible with a few sidelobes due to dynamic range problems. Some
indication of the two short symmetric jets is present and the eastern hot spot
is marginally resolved while the western one is completely resolved. At the
highest resolution (uniformly weighted map, HPBW = 130 mas) only a slightly
resolved core ($\sim 30 mas$ in size) is visible.

\begin{figure}
\figurenum{4}
\plotfiddle{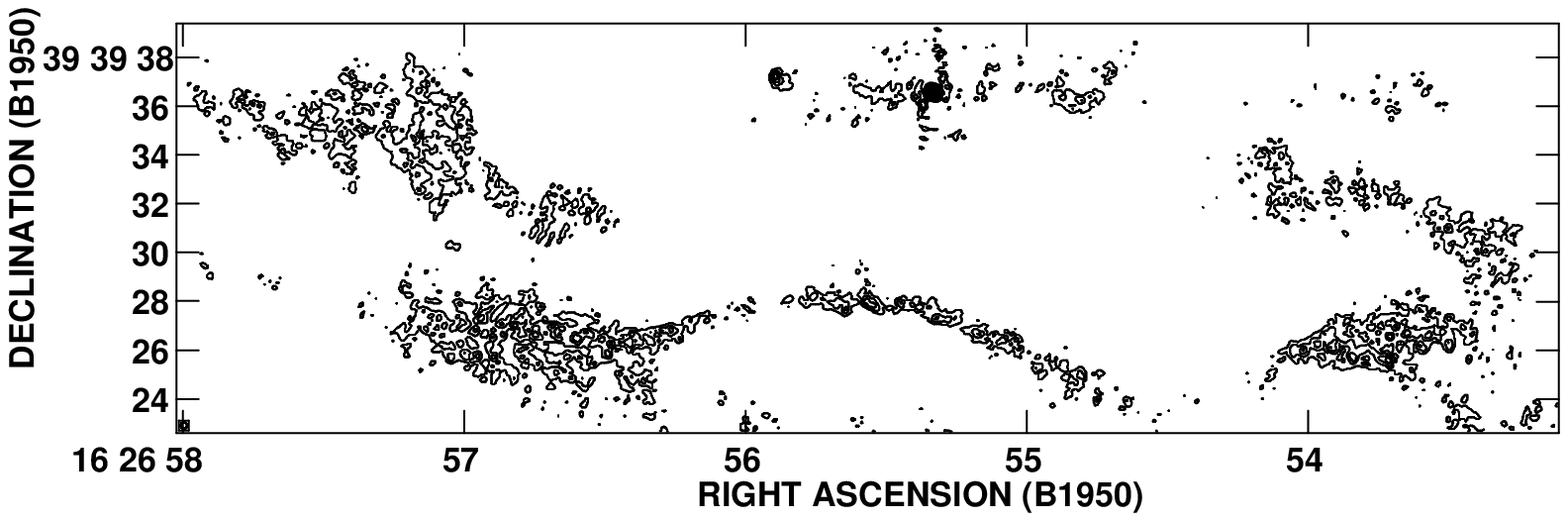}{3.5truecm}{0}{60}{60}{-200}{-200}
\caption{
Isocontour map of 3C338 at 1.7 GHz obtained with the MERLIN
array. The HPBW is 250 mas and the noise level is 0.13 mJy/beam.
Contour levels are: 0.25 0.5 0.75 1 3 5 10 30 50 100 150 mJy/beam.}
\end{figure}

No polarized flux has been detected in the MERLIN data at a level of 0.1 
mJy/beam. 

\subsection{VLBI Data}

Table 2 summarizes the VLBI observations which are presented in the
following sub-sections. The noise level and angular resolution (HPBW) are given
for natural weighted maps.

\begin{table}
\caption{VLBI maps \label{t2}}
\begin{flushleft}
\begin{tabular}{lllll}
\tableline
\noalign{\smallskip}
Frequency & Array & Obs. data & HPBW & noise \\
GHz       &       &           &  mas & mJy/b \\
\noalign{\smallskip}
\tableline
\noalign{\smallskip}
1.7       & Global& Jun91 & 8.2$\times$3.9& 0.06 \\
5.0       & Global& Sep89 & 3.2$\times$3.2& 0.3  \\
5.0       & VLBA+Y& Nov94 & 3.2$\times$3.2& 0.2  \\
5.0       & VLBA+Y& Sep95 & 3.5$\times$2.7& 0.25 \\
8.4       & Global& Mar91 & 2.0$\times$1.0& 0.09 \\
8.4       & VLBA+Y& Nov94 & 2.2$\times$2.2& 0.15 \\
8.4       & VLBA+Y& Sep95 & 2.1$\times$1.9& 0.1  \\
\noalign{\smallskip}
\tableline
\noalign{\smallskip}
\label{t2}
\end{tabular}
\end{flushleft}
\end{table}

\subsubsection{Data at 1.7 GHz}

We observed 3C338 at 1.7 GHz on 1991 June for 12 hours with the MK3 mode 
B recording system (28 MHz bandwidth) with the following global array: Bonn,
Jodrell MK1, Medicina, Onsala, WSRT, Green Bank, Haystack, VLBA-Owens Valley
and Pie Town. Standard amplitude
calibration was done using the system temperature method in AIPS.
Data were then globally fringe fitted and self-calibrated in the standard way.
We made 6 iterations of phase self-calibration and 1 phase+gain to produce the
final map shown in Fig. 5. The source shows a central 
component that we identify
with the nuclear source and a symmetric structure extended about 30 mas to the
east and 20 mas to west. The faint, more
diffuse emission visible to the west at 40 mas from the core is probably real. 
The size and position
angle of the symmetric structure is in agreement with the ``core'' parameters
derived from the full resolution MERLIN map. 

\begin{figure}
\figurenum{5}
\plotfiddle{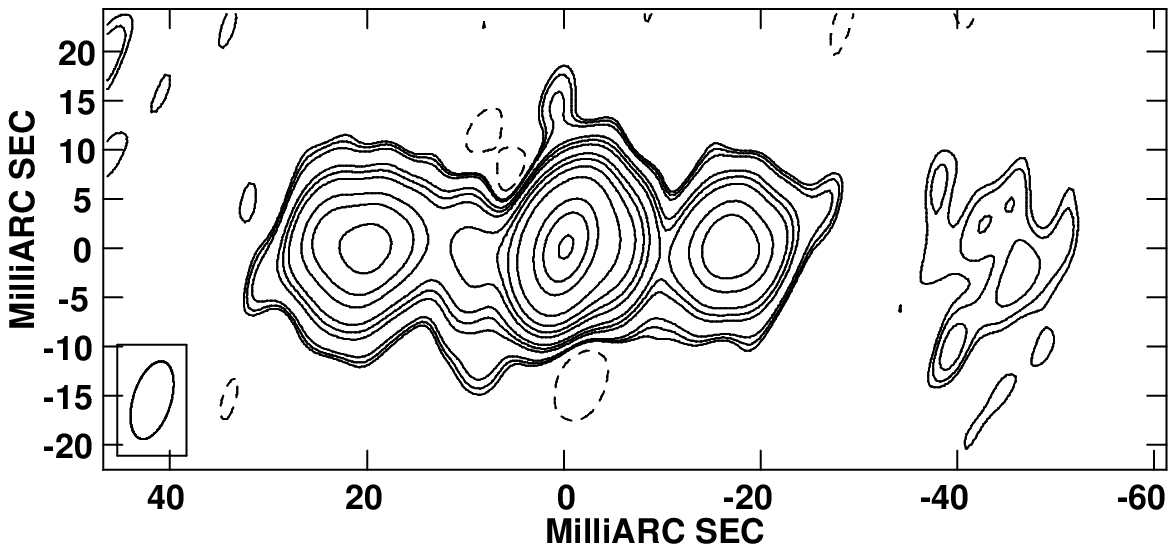}{4truecm}{0}{70}{70}{-220}{-220}
\caption{
VLBI map of 3C338 at 1.7 GHz. The HPBW is 8.2 $\times$ 3.9 mas
in PA -17$^{\circ}$. The noise level is 0.06 mJy/beam.
Contour levels are: -0.15 0.12 0.15 0.2 0.5 0.7 1 2 3 5 10 30 50 80 mJy/beam.}
\end{figure}

\subsubsection{ Data at 5 GHz}

At 5 GHz 3C338 was mapped by us for the first time in 1989 September (see
Feretti et al., 1993), and observed again on 1994 November and
1995 September with the full VLBA + the phased VLA. The
observations were carried out switching every 30 minutes from 5 to 8.4 GHz
obtaining two maps at two observing frequencies at the same epoch and with
good uv-coverage. The data have been correlated in Socorro.
The parsec scale structure shows a central dominant
feature (the core emission) and two symmetric jets. The eastern jet is
slightly stronger and longer (Fig. 6).
\begin{figure}
\figurenum{6}
\plotfiddle{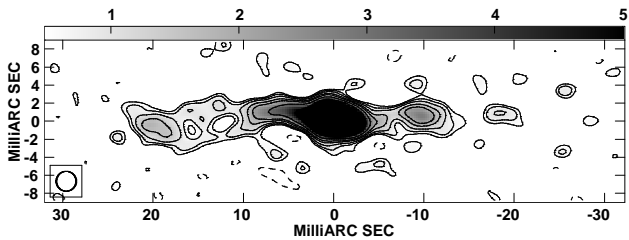}{3.0truecm}{0}{60}{60}{-180}{-190}
\caption{
Isocontour map of 3C338 obtained with VLBA + Y27 at 5 GHz
on November 1994. The HPBW is 2.2 mas and the noise level is 0.2 mJy/beam
Contour levels are:-0.5 0.5 0.7 1 1.5 2 3 5 10 30 40 mJy/beam}
\end{figure}
The eastern jet shows a couple of low brightness regions in its center
suggesting it could be limb-brightened. Unfortunately, the 1995 data were 
seriously affected by the lack of data on the calibrator source
due to technical problems and by the failure of some telescopes. 
Therefore these data produced a
low quality map which does not add any useful detailed information about the
source structure. 

\subsubsection{Data at 8.4 GHz}

We observed 3C338 at 8.4 GHz on March 1991 with the following array: Bonn,
Medicina, Noto, Onsala, Green Bank, Haystack, OVRO, VLA (phased array),
VLBA-Pie Town and Kitt Peak. The data have been correlated in Bonn.
Second and third epoch maps were obtained with the full VLBA + VLA
phased array at the same time as the 5 GHz maps (see above). The 3 epoch maps
are given in Fig. 7. At 8.4 GHz the
source is symmetric, but the eastern jet is somewhat longer and brighter
than the western jet. The central component is dominant in all maps, however
it shows a clear change of structure in the 3 epochs.

\begin{figure*}
\figurenum{7}
\plotfiddle{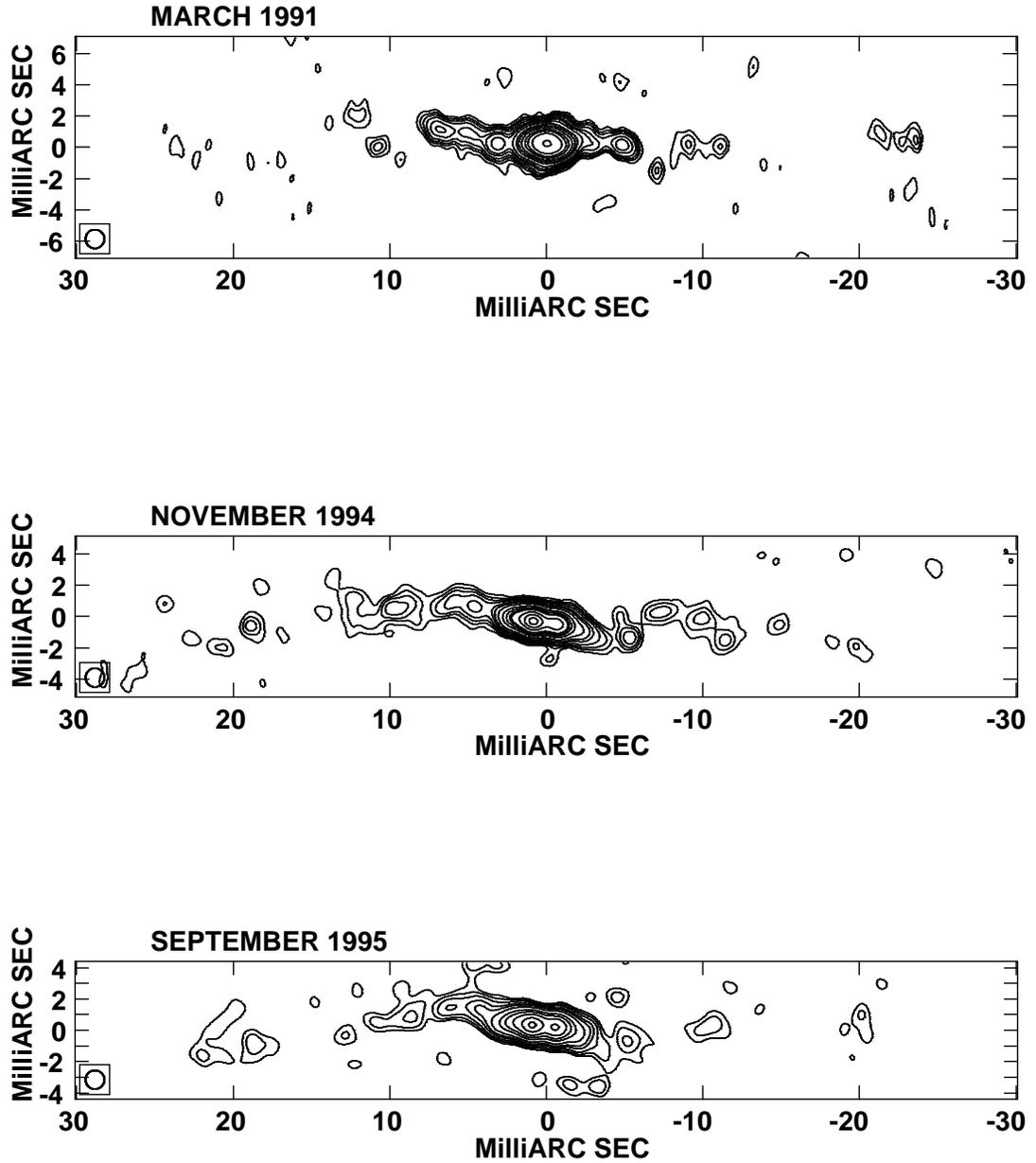}{20truecm}{0}{110}{110}{-350}{-200}
\caption{
Multi epoch maps of 3C338 at 8.4 GHz. The peak flux is 64.0
mJy/beam for the first epoch map (top), 27.1 mJy/beam for the second epoch map
(middle) and 21.0 for the third epoch map (bottom). For all the maps,
the HPBW is 1.2 mas and contour levels are: 0.5 0.7 1 1.5 2 3 5 7 10 15 20 30
60 mJy/beam. }
\end{figure*}

\begin{figure*}
\figurenum{8}
\plotfiddle{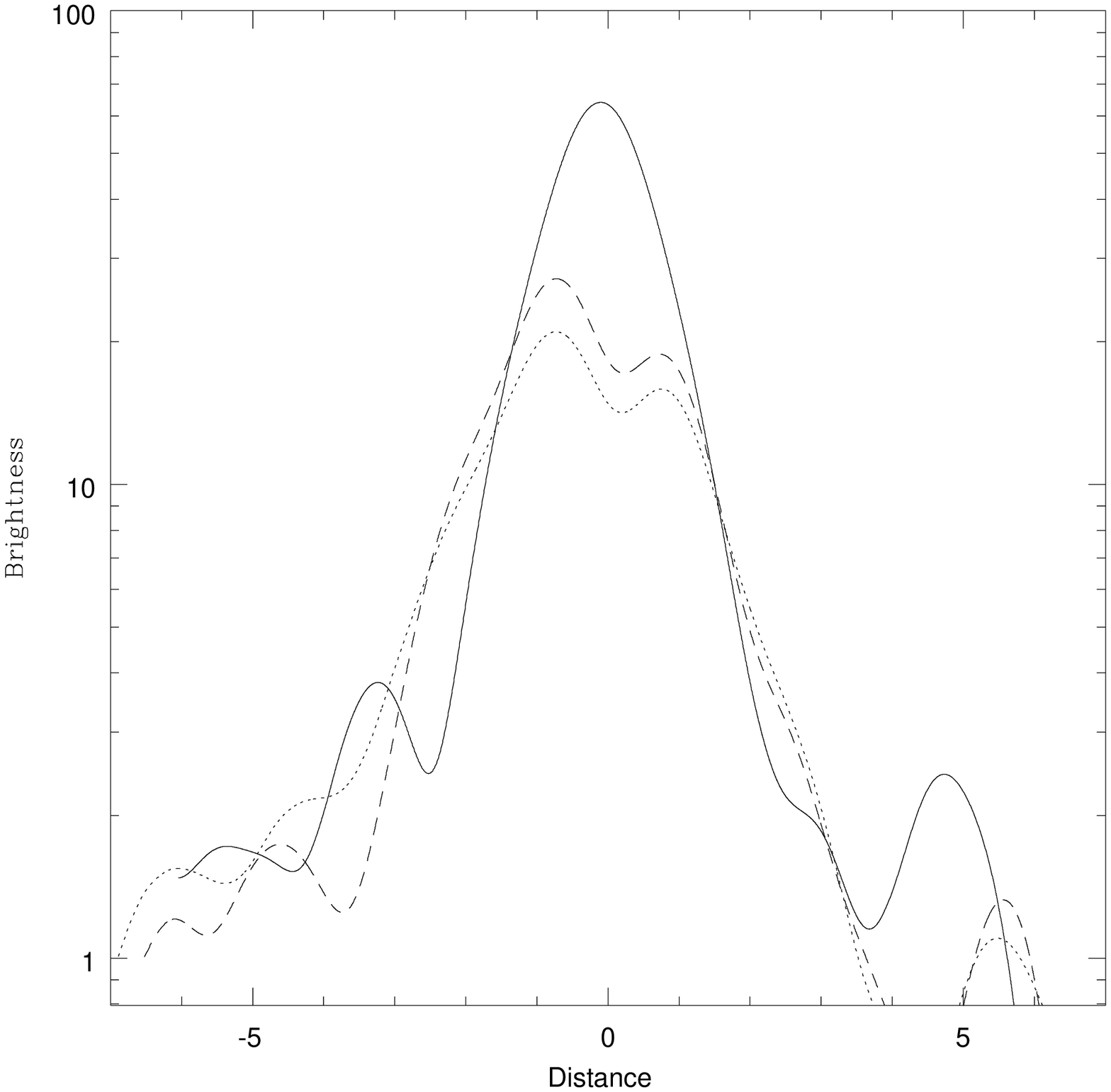}{8truecm}{0}{45}{45}{-120}{-72}
\caption{
Overlay of the brightness of the 3 epochs, assuming that the core
is between the two component detected in the second and third epoch.
The continuous, dashed and dotted lines refer to the first, second and third
epoch, respectively.}
\end{figure*}

\section{Discussion}

\subsection{Changes in the pc scale structure and proper motion}

With the VLBI observations at our disposal, we
performed a multi-epoch morphological comparison of 3C338. 

At 8.4 GHz, the morphological change between different epochs is quite 
evident (Fig. 7). In particular, the map of the first epoch is dominated 
by a single central component of high brightness,
while the more recent observations 
show two resolved components in the core region.
Moreover, there are slight changes in the location and position angle of the 
blobs.
The maps at 5 GHz are consistent with this change of structure, although
their resolution is lower.

To better enhance the structural variations, we produced slices
of the brightness along the ridge of maximum brightness, in the 
innermost region, i.e. where the brightness is higher (see Fig. 8).
The alignment of the first and second epoch map is not unique,
therefore we distinguish the following possible cases.

{\bf Case 1.} The central peak in the first epoch map coincides with the
true core and the two central components detected in the
second and third epoch are new ejecta, separated by 1.5 mas, with
the true core in between (see Fig. 8).
The central structure detected in the first epoch is slightly resolved
($\sim$ 1.5 mas) and could be produced by a double source,
with point-like identical components, of separation $\sim$1.25 mas.
Therefore, the new ejecta
have moved of at least 0.12 mas between the first and second epoch.
A larger proper motion is seen in three outer blobs.
Since the blobs are weak, they could be affected by positional uncertainties
related with the application of the clean algorithm. On average, we estimate 
that each blob has moved away from the core of about
1.1-1.2 mas between the two epochs. This leads to an apparent blob velocity
$\beta_{app}$= 0.41h$^{-1}$ - 0.45h$^{-1}$, (h = H$_0$/100) 
which corresponds to a true velocity $\beta >$ 0.38h$^{-1}$ - 0.41h$^{-1}$.
The apparent blob  velocity is consistent with the lack of proper
motion between the second and third epoch. In fact, the implied displacement
of the blobs between the two observations should be 0.23--0.25 mas.
The apparent velocity of the two symmetric innermost blobs is lower. Features
moving at different velocities, as well as moving and
stationary components are common in extragalactic radio
sources (see e.g. Zensus et al., 1995). We will consider here as 3C338
proper motion the faster one, derived from the outer components.

{\bf Case 2.} The easternmost strong peak detected  in the second and
third epoch contains the true core, and coincides with the peak of the
first epoch (Fig. 8). The motion of the blobs is very similar to that
obtained in the previous case, but the emission is asymmetric and the blob 
motion too. A flip-flop mechanism could explain the two-sided morphology.

{\bf Case 3.} The westernmost strong peak detected in the second and
third epoch is the true core, while the easternmost one is a new
component. In this case, the new component shows an apparent velocity
$\beta$=0.63h$^{-1}$, while the outer blobs seem 
stationary between the first and
second epoch. An upper limit to their velocity of 0.1 c is obtained.
If this is the case, the source is very peculiar, with a large velocity
decrease of the blobs shortly after their ejection. Alternatively they could be
stationary components due to a bend in the jet, although no large bend is 
present at pc and kpc resolution in this source. As in case 2, a flip-flop
mechanism should be involved.

We consider the first possibility as the most reliable, because
the two-sided ejection is what we expect given the large scale symmetry and 
because the two components show a similar behaviour in the flux density
variation between the second and third epoch. Therefore,
hereafter we will discuss only the case of the symmetric emission.

\subsection{ Flux variability and spectrum}

The morphological change discussed in the previous subsections is related to 
the variability of arcsecond core flux density.
The first epoch map was made in a
period of maximum flux density of the arcsecond core, while the second and
third epochs are in a low flux density stage. 

A comparison of the images obtained in November 1994 at 5 and 8.4 GHz
indicates that the spectral index of the double source at the center
is 0.4. The flux density of both components decreases 
from the second to the third
epoch, by almost the same amount (15-20\%). This similar behaviour
reinforces the interpretation that the two components are two ejected blobs,
moving away from the core. 
Their flux density variation could be related to
adiabatic expansion during the propagation.
We also note that the strongest peak is the eastern one, i.e. on the
same side of the main jet (Sects. 2.3.2 and 2.3.3).

We can therefore suggest that the high state of flux density in the
arcsec core (see Fig. 1 and Table 1) is related to the emission of new 
components, while the following
flux density decrease is related to the propagation and expansion of these
components. The emission of the two central components visible in the 
second epoch map would be related to the high arcsecond core flux 
density measured in 1991.

\subsection{Source orientation with respect to the line of sight}

In Giovannini et al. (1994) and Lara et al. (1997), the jet velocity 
and orientation were derived from observational constraints.
Here we obtain the jet orientation and velocity in the light of the 
better estimate of the 
jet to counter-jet brightness ratio and of the proper motion measure. 
The eastern jet appears to be the main one  as  it 
shows a higher brightness, it
is visible to a larger distance from the
core and it is on the same side of the more compact "VLA" hot spot
(see Fig. 5 and  Sects. 2.3.2 and 2.3.3). We 
have measured a jet/counter--jet (j/cj) ratio R $\sim$ 1.4. 
Assuming a jet spectral index = 0.5, with the relation
\begin{eqnarray}
  R = \left( {1 + \beta cos \theta \over 1 - \beta cos \theta} \right) ^{2 +
\alpha}  
\end{eqnarray}
we obtain $\beta$~cos$\theta$ $\sim$ 0.07. 
\goodbreak
\parindent 0pt
From equation (1) and from the relation between $\beta$ and $\beta_{app}$:
\begin{equation}
 \beta = {\beta_{app} \over \beta_{app}cos\theta + sin\theta} 
\end{equation}
we have
\begin{equation}
  \tan \theta = {2 \beta_{app} \over R^{1/(2 + \alpha)} -1 } 
\end{equation}
Assuming $\beta_{app}$ = 0.43 (see Sect. 3.2) we derive that the parsec scale
structure of 3C338 is at $\sim$
80$^{\circ}$ with respect to the line of sight, the jet velocity is
$\beta \sim$ 0.40, and the corresponding Doppler factor
$\delta$ is 0.99. A value of $\delta$ close to 1 means that the two jets are 
not strongly de-boosted.

Of course this result implies a bulk jet velocity of the same 
order of the 
velocity of the moving knots (pattern velocity).
We cannot exclude a jet with a bulk velocity higher than the pattern one,
however 
a high jet velocity implies in any case an angle near to the plane of the
sky since we have symmetric jets. A jet velocity of 0.9c is possible at an 
angle of 86$^{\circ}$ with
respect to the line of sight, but it implies $\delta$ = 0.47 i.e. a strong 
de-boosting
of the jet brightness. Therefore, either the 3C338 jets are very
peculiar with a 
very high intrinsic brightness or $\delta$ has to be $\sim$ 1. 
Another possibility is to have jets slower than the moving knots.
In this case no constraint can be
given to their orientation with respect to the line of sight. In fact
large as well as small angles are possible 
for jets with a low bulk velocity (see Giovannini et al., 1994).

In the light of the Ghisellini et al. (1993) result and of the 
general agreement
in literature between the bulk and pattern velocity for superluminal sources
we will consider here that in 3C338 the jet velocity is very similar to the
velocity derived from the moving knots. 

We have analyzed how the present results are affected by the choice of the
Hubble constant. If H$_0$ is 50 km sec$^{-1}$ Mpc$^{-1}$ the
jet velocity becomes 0.8c and $\theta$ $\sim$ 85$^{\circ}$. In this case
the Doppler factor is 0.64 and the two sided jets are de-boosted.
The intrinsic flux density of the parsec scale structure is a factor
$\sim$ 4.8 higher than the measured
one and the radio power of the parsec scale structure
at 5 GHz becomes $\sim$ 10$^{24.5}$ W/Hz.
Therefore, the present data are
consistent with both H$_0$ = 100 and H$_0$ = 50 km sec$^{-1}$ Mpc$^{-1}$
and only allow us to derive the H$_0$ lower limit: H$_0$ $\geq$ 40 km sec$^{-1}$
Mpc$^{-1}$.

We have also checked whether the previously obtained velocity and jet
orientation can account for the different jet lengths
visible in our VLA maps (see Sect. 2.1.2). If we assume that this difference is
due to the different jet direction (the eastern jet is approaching us
while the western one is receding), the ratio of projected distances from
the core is given by:
\begin{eqnarray}
 {Length_{approach} \over Length_{reced}} = 
 {1 + \beta cos \theta \over 1 - \beta cos \theta } 
\end{eqnarray}
Assuming a constant jet velocity, 
and an arm length ratio = 1.07, we derive
$\beta cos\theta$ $\sim$ 0.03 in good agreement with the value from the j/cj
ratio. The lower value found from the arm length ratio 
is expected for a jet decelerating at increasing distance from the core.

\subsection{Source structure}

The overall structure of 3C338 consists of two features, with very different
properties:  an active region, 
which includes the core, two symmetric jets and two faint hot-spots 
at the jet ends, 
and a diffuse region, displaced to the south, 
showing a jet like filament and low-brightness extended emission. 
There is no visible connection between the symmetric jets
and the diffuse feature, which is characterized by very steep spectrum, and
is then likely to be an old relic radio emission.

The properties of the active structure (as flux variability, proper
motion and so on; see previous sections) support the model of Burns et al.
(1983) that this is a re-born young radio source.
If the hot spot advance velocity in the surrounding medium is
$\sim$ 0.02c (the 'fast' model in Readhead et al., 1996),  
the time necessary to reach the present size is 
5 $\times$ 10$^5$ yrs. This value  becomes 2.5 $\times$ 10$^6$ yrs 
if the 'slow' model (Readhead et al. 1996) is assumed.  
We note that this structure fits very well in the class of {\it medium size
symmetric objects} (MSOs) defined by 
Readhead et al., (1996) and Fanti et al., (1995). The only discrepancy is 
that 3C338 is a nearby low power source while the Compact Symmetric 
Objects (CSOs) and the MSOs discussed by Readhead et al., (1996) and 
Fanti et al., (1995) are high redshift,
high power objects. In the evolutionary scenario, these high power compact
sources expand into FR II sources and are therefore considered the likely 
progenitors of this class of sources. The young radio source in 3C338 could be 
a low power MSO as 4C31.04 (Cotton et al., 1995) which we expect to grow into 
an extended FR I source. 
The two faint hot spots at the jet
ends  could represent the regions of interaction between the jets and the
dense surrounding medium. 
The existence  of {\it recurrent sources} is also
suggested by O'Dea (1997) as a possible scenario to explain the fraction of
Gigahertz Peaked Sources (GPS) with extended emission. 
In fact, if the new phase of activity
begins while the extended relic is still visible, this should provide a compact
young source with an old extended emission. Unlike GPSs (O'Dea, 1996), 3C 338
is embedded in a strong cooling flow, which prevents adiabatic
losses. Therefore the relic emission is still visible when the youngest
emission has evolved into a MSO source.

The parsec scale jets are bright and well collimated 
up to $\sim$ 30 mas ($\sim$ 12 pc) from the core. This is consistent with the 
extent of the  nuclear source in the MERLIN map, and 
the structure in the VLBI 18 cm map. Beyond that distance, 
the jets become weaker, and are visible only in  the
VLA image and marginally detected in the low resolution MERLIN map. 
In analogy with the properties of 3C264 (Baum et al. 1997), a strong
jet deceleration coupled with a rapid expansion which produces a strong 
brightness decrease because of adiabatic losses, could be present.
The {\it decollimation} region occurs very close to the 
nuclear source ($\sim$ 12 pc) while in 3C264 the jet is well collimated up to 
$\gtrsim$ 200 pc. This difference can be related to the low power {\it young} 
radio emission of 3C338 or to  interaction with the outer medium.
Unfortunately, because of the presence of many blobs/shocks
in the 3C338 jets, it is inconclusive to model their brightness and opening 
angle, and to derive the trend of velocity (see Baum et al. 1997).
Arc second scale jets visible in the VLA maps (see Fig. 3)
seem to be intrinsically connected to the parsec scale jets. They are well
aligned and have similar properties.
In fact,  on both
scales, the eastern jet is slightly dominant over the western one. 

It is important to note that, despite of the expected orbital motion 
of the multiple optical nuclei (Burns et al., 1983), the radio jet direction 
is stable in time: the PA of the restarted jet is very close to the PA of
the jet-like filament in the relic  emission and
 at the same PA of the extended structure.

Owen et al. (1997) found an asymmetric X-ray
emission around 3C338. Since the X-ray emission is extended in the jet
direction, they suggested that the radio jet is transferring momentum
to the X-ray gas, pushing it out. This requires a
radio jet direction constant in time. The authors also
suggest that the arcsecond scale jet becomes unstable and disrupts
severely at about the present position,
because of the interaction with the surrounding medium.
Their derived age for the jet structure
is $\sim$ 17 $\times$ 10$^6$ yrs, larger than that derived by us, using the 
Readhead et al. (1996) model, but not in conflict.
A larger source age implies a jet advance speed lower than that 
derived by Readhead et al. (1996) for MSOs and this lower velocity could be
once again related
to the low power of 3C338 with respect to classical MSOs and 
to the dense 3C338 environment.

\section{Conclusions}

We have presented here new VLA, MERLIN and VLBI maps of the radio 
galaxy 3C 338. Moreover we have collected all the available data on its 
arcsecond core flux density. The source shows a strong flux density 
variability with at least
two epochs of major activity in the range 1974 -- 1995.

The VLA and MERLIN images show the presence of a core emission with two-sided
jets disconnected from the large scale emission. The VLBI data confirm the
existence of symmetric pc scale jets. The orientation of this structure
appears to be very constant in time despite of the complex dynamic conditions
present in the 3C338 central regions.

Comparing maps obtained at different 
epochs, a change in the pc scale morphology is well evident, and it is
probably correlated 
with the arcsecond core flux density variability.
The structural changes suggest the presence of a proper motion with $\beta 
\sim$ 0.4h$^{-1}$ on both sides of the core. This symmetric motion allow us
to constrain the Hubble constant to H$_0$ $\geq$ 40 km sec$^{-1}$ Mpc$^{-1}$.

These properties suggest that the extended emission in 3C 338 is a relic 
structure not related to the present nuclear activity, whose age is
comparable with that of the high power and distant MSOs discussed by
Fanti et al., (1995) and Readhead et al., (1996). The low power of this young 
emission is in agreement with its evolution in a FR I radio source while
more distant MSOs are expected to evolve in FR II sources.

\acknowledgments

We thank Dr. R. Fanti for useful suggestions, Dr. M. Bondi, Dr. D. Dallacasa 
and Dr. C. Fanti for a critical reading of the manuscript.
We thank the staffs at the telescopes for their assistance with the
observations and the Bonn and Socorro correlator people for the absentee
correlation of the data.
The National Radio Astronomy Observatory
is operated by Associated Universities, Inc., under contract with the 
National Science Foundation.


\newpage
\begin{thebibliography}{}

\bibitem{}
Baum, S.A., O'Dea, C.P., Giovannini G., Biretta J., Cotton, W.D., de Koff S., 
Feretti 
L., Golombek D., Lara L., Macchetto F.D., Miley G.K., Sparks W.B.,
Venturi T., Komissarov S.: 1997 \apj 483, 178.
\bibitem{}
Burbidge, E.M.: 1962 \apj 136, 1134
\bibitem{}
Burns, J.O., Schwendeman, E., White, R.A.: 1983, \apj 271, 575
\bibitem{}
Cotton, W.D., Feretti, L., Giovannini, G., Venturi, T., Lara, L., Marcaide,
J., Wehrle, A.E.: 1995 \apj 452, 605
\bibitem{}
Ekers, R.D., Fanti, R., Miley, G.K.: 1983, A.A. 120, 297
\bibitem{}
Fanaroff, B.L., Riley J.M.: 1974, \mnras  167, 31
\bibitem{}
Fanti C., Fanti R., Dallacasa D., Schilizzi R.T., Spencer R.E., Stanghellini 
C.: 1995 \aap 302, 317
\bibitem{}
Feretti L., Comoretto G., Giovannini G., Venturi T., Wehrle A.E.: 1993, 
\apj 408, 446
\bibitem{}
Fisher, D., Illingworth, G., Franx, M.: 1995 \apj 438, 539
\bibitem{}
Gavazzi, G., Boselli, A., Carrasco, L.: 1995 \aaps 112, 257
\bibitem{}
Ge, J., Owen, F.N.: 1994, \aj 108, 1523
\bibitem{}
Ghisellini, G., Padovani, P., Celotti, A., Maraschi, L.: 1993 \apj 407, 65
\bibitem{}
Giovannini, G., Feretti, L., Comoretto, G.: 1990 \apj 358, 159
\bibitem{}
Giovannini, G., Feretti L., Venturi T., Lara L., Marcaide J., Rioja M.,
Spangler S.R., Wehrle A.E.: 1994 \apj 435, 116
\bibitem{}
Lara, L., Cotton, W.D., Feretti, L., Giovannini, G., Venturi, T., Marcaide, 
J.M.: 1997 \apj 474, 179
\bibitem{}
Lucey, J.R., Gray, P.M., Carter, D., Terlevich, R.J.: 1991 \mnras 248, 804
\bibitem{}
Minkowski R.: 1961 \aj 66, 558
\bibitem{}
O'Dea, C.P.: 1996 Proceedings of the Second Workshop on Gigahertz Peaked
Spectrum and Compact Steep Spectrum Radio Sources, I.A.G. Snellen, R.T.
Schilizzi, H.J.A. Roettgering and M.N. Bremer eds, p. 142
\bibitem{}
O'Dea, C.P.: 1977 \pasp in press
\bibitem{}
Owen, F.N., Eilek, J.A.: 1997 \apj in press
\bibitem{}
Parma, P., de Ruiter, H.R., Fanti, C., Fanti, R.: 1986, \aaps 64, 135
\bibitem{}
Readhead, A.C.S., Taylor G.B., Xu W., Pearson T.J.: 1996 \apj 460, 612
\bibitem{}
De Ruiter, H.R., Parma, P., Fanti, C., Fanti, R.: 1986, \aaps 65, 111
\bibitem{}
Zabludoff, A.I., Geller, M.J., Huchra, J.P., Vogeley, M.S.: 1993 \aj 106, 1273
\bibitem{}
Zensus J.A. Krichbaum, T.P., Lobanov, A.P.: 1995 Proc. Natl. Acad. Sci. USA
92, 11348

\end{thebibliography}
\end{document}